\documentclass{sf2a-conf2026}
\usepackage{graphicx}
\usepackage{hyperref}
\usepackage[]{natbib}  
\usepackage{epstopdf}
\usepackage{amsmath,amssymb}
\usepackage[export]{adjustbox}

\def\BibTeX{{\rm B\kern-.05em{\sc i\kern-.025em b}\kern-.08em
    T\kern-.1667em\lower.7ex\hbox{E}\kern-.125emX}}
\bibpunct{(}{)}{;}{a}{}{,}  

\begin{document}

\TitreGlobal{SF2A 2026}

\title{AGN winds and outflows: \\ from accretion-disc scales to the host galaxy}

\runningtitle{AGN winds and outflows: from accretion-disc scales to the host galaxy}

\author{D. Porquet}\address{Aix Marseille Univ, CNRS, CNES, LAM, Marseille, France}
\setcounter{page}{237}


\maketitle

\begin{abstract}
Ultra-fast outflows (UFOs), launched from the inner accretion disc in active galactic nuclei (AGN), are among the prime candidates for driving AGN feedback in their host galaxies. Determining their physical properties and energetics is therefore essential for assessing their impact on galaxy evolution. This review highlights recent observational progress, with particular emphasis on the groundbreaking insights enabled by high-resolution X-ray spectroscopy. The first \textit{X-Ray Imaging and Spectroscopy Mission} (XRISM)/\textit{Resolve} observations of AGN winds have revealed that the innermost UFOs are kinematically structured and inhomogeneous. In the archetypal quasar PDS\,456, these observations have led to a substantial upward revision of the inferred mass outflow rate, momentum flux and kinetic power. Complementary UV, optical, IR and mm observations trace the ionised, atomic and molecular phases of AGN-driven outflows from parsec to galactic scales. However, the physical coupling between the accretion-disc wind and the galactic  outflows remains one of the major unresolved questions in AGN feedback. 
Building on the pioneering \textit{XRISM} results, \textit{NewAthena’s X-ray Integral Field Unit} (X-IFU) will characterise the physical properties and energetics of accretion-disc winds with unprecedented precision from the local Universe up to cosmic noon, while complementary multi-wavelength observations with facilities such as the \textit{Extremely Large Telescope} (ELT), \textit{James Webb Space Telescope} (JWST), \textit{Atacama Large Millimeter/submillimeter Array} (ALMA), and \textit{Square Kilometre Array (SKA)} will determine whether and how their energy and momentum are transferred to the host galaxy.
\end{abstract}



\section{Introduction}

Supermassive black holes (SMBHs) are now known to reside at the centres of most massive galaxies. Although their spheres of gravitational influence are much smaller than their host galaxies, numerous studies have revealed correlations between the SMBH mass and the stellar velocity dispersion, luminosity, and stellar mass of the host-galaxy bulge (e.g. \citealt{Ferrarese2000,Gebhardt2000,Kormendy2013}). These scaling relations suggest that SMBHs and their host bulges co-evolve. Understanding this connection requires characterising both the feeding of the SMBH through gas accretion and AGN feedback. Accretion powers radiation, winds and jets, which can transfer energy and momentum to the surrounding gas. By heating, expelling or redistributing this gas, AGN feedback may regulate black-hole growth and star formation, thereby contributing to the observed scaling relations (e.g. \citealt{DiMatteo2005,Fabian2012,King2015}).\\

This review focuses on AGN winds and outflows, from sub-parsec accretion-disc scales to galaxy-scale outflows extending over tens of kiloparsecs. Their highly ionised, lower-ionisation, neutral and molecular components are traced through complementary X-ray, UV, optical, IR and mm observations. Their \textit{multi-scale}, \textit{multiphase}, and \textit{time-variable} nature presents major challenges for both observations and theoretical modelling (e.g. \citealt{Cicone2018,Gaspari2020}). 
Section~\ref{sec:X-rays} reviews the main properties of X-ray ionised winds. 
Section~\ref{sec:X-ray spectro} presents recent results from the \textit{Resolve} X-ray microcalorimeter onboard the \textit{X-Ray Imaging and Spectroscopy Mission} (XRISM), with particular emphasis on the luminous, near-Eddington quasar PDS\,456, and discusses prospects with ESA’s future X-ray observatory, \textit{NewAthena}. 
Section~\ref{sec:multi-wav} examines the multi-wavelength constraints on the connection between accretion-disc winds and galaxy-scale outflows.
Finally, Section~\ref{sec:conclusion} summarises the main conclusions and the role of current and future facilities in assessing the contribution of accretion-disc winds to AGN feedback.

\section{The X-ray view of ionised AGN winds}\label{sec:X-rays}

X-ray spectroscopy probes the innermost regions of AGN and the circumnuclear gas illuminated by the primary continuum. 
When ionised gas intercepts our line of sight to the X-ray source, it produces absorption features whose energies, profiles and relative strengths constrain its kinematics, ionisation state and column density through photoionisation modelling.
X-ray ionised outflows have traditionally been classified into two main categories: \textit{warm absorbers} (WAs), detected primarily in the soft X-ray band ($\lesssim$2--3\,keV), and \textit{ultra-fast outflows} (UFOs), identified mainly through blueshifted Fe\,{\sc xxv} and Fe\,{\sc xxvi} absorption lines, generally detected above their rest energies of 6.70 and 6.97\,keV, respectively.\\

\textit{WAs} were extensively studied with the \textit{Advanced Satellite for Cosmology and Astrophysics} (ASCA) in the 1990s and subsequently characterised in greater detail with the high-resolution grating spectrometers onboard \textit{Chandra} and \textit{XMM-Newton}, both launched in 1999. WAs produce absorption lines from H-like and He-like ions of C to S, together with Fe M-shell unresolved transition arrays, tracing gas with typical hydrogen column densities $N_{\rm H}$$\lesssim$10$^{22}$\,cm$^{-2}$, ionisation parameters $\log\xi$$\lesssim$3, and outflow velocities typically below $\sim$10$^3$\,km\,s$^{-1}$. Here, $\xi$=$L_{\rm ion}$/$(n_{\rm H}r^2)$, where $L_{\rm ion}$ is the 1--1000\,Ryd ionising luminosity, $n_{\rm H}$ the hydrogen number density and $r$ the distance from the ionising source; $\xi$ is expressed in erg\,cm\,s$^{-1}$. WAs are detected in approximately 50--65\% of nearby Seyfert\,1 galaxies \citep[e.g.][]{Porquet2004,Patrick2012,Laha2014}. Their inferred distances range from sub-parsec scales comparable to the broad-line region (BLR) to the dusty torus and narrow-line region (NLR), with some components located hundreds of parsecs or even kiloparsecs from the nucleus \citep[e.g.][]{Porquet1999,Kaastra2012,Ebrero2016}.\\

Classical \textit{UFOs} are characterised by outflow velocities of $\sim$0.03--0.5c, large column densities ($N_{\rm H}$$\gtrsim$10$^{22}$\,cm$^{-2}$) and high ionisation parameters ($\log\xi$$\gtrsim$3). Systematic searches yield detection fractions of about one third to one half in nearby AGN samples \citep[e.g.][]{Tombesi2010,Patrick2012,Gianolli2024}. 
UFOs are also detected beyond the local Universe, from intermediate-redshift AGN at $z\simeq0.1$--$0.4$ \citep{Matzeu2023} to lensed and non-lensed quasars spanning $z\simeq1.4$--$3.9$ \citep{Chartas2021,Lanzuisi2026}. 
Their inferred mass outflow rates and velocities can yield kinetic powers comparable to or exceeding the $\sim$0.5--5\% $L_{\rm bol}$ thresholds invoked in some AGN feedback models \citep[e.g.][]{DiMatteo2005,Hopkins2010}. 
UFOs are therefore considered strong candidates for driving galaxy-scale feedback.\\

Although the distinction between WAs and UFOs remains observationally useful, ionised AGN outflows span a broader range of properties than this classification implies. Low-ionisation UFOs combine the high velocities of classical UFOs with ionisation parameters and column densities similar to those of WAs \citep[e.g.][]{Yamada2024}. 
In the type\,1 AGN PG\,1114+445, \citet{Serafinelli2019} interpreted a low-ionisation, high-velocity absorber as ambient gas entrained and accelerated by an inner UFO and referred to this component as an \textit{entrained UFO}.
If confirmed in larger samples, such entrainment would provide an observational link between accretion-disc winds and more extended ionised gas. In addition, \textit{ultra-thick WAs} (UTWAs) have WA-like ionisation parameters but column densities comparable to those of classical UFOs. Some may represent lower-ionisation phases associated with accretion-disc winds \citep{Middei2026}. These absorbers indicate that different ionisation and velocity components may be physically connected, although a common origin for all WAs and UFOs remains unestablished.\\

Testing these possible connections requires high-resolution X-ray measurements of the ionisation and velocity structure of the absorbing gas. The \textit{Resolve} X-ray microcalorimeter onboard \textit{XRISM}  \citep{Tashiro2025} now allows the Fe\,K absorption structure of highly ionised UFOs to be resolved in substantially greater detail, as discussed in the following section.

\section{The X-ray microcalorimeter revolution in the study of AGN accretion-disc winds}\label{sec:X-ray spectro}

\begin{figure}[t]
\centering
\begin{tabular}{cc}
\includegraphics[width=0.435\textwidth,valign=t,clip]{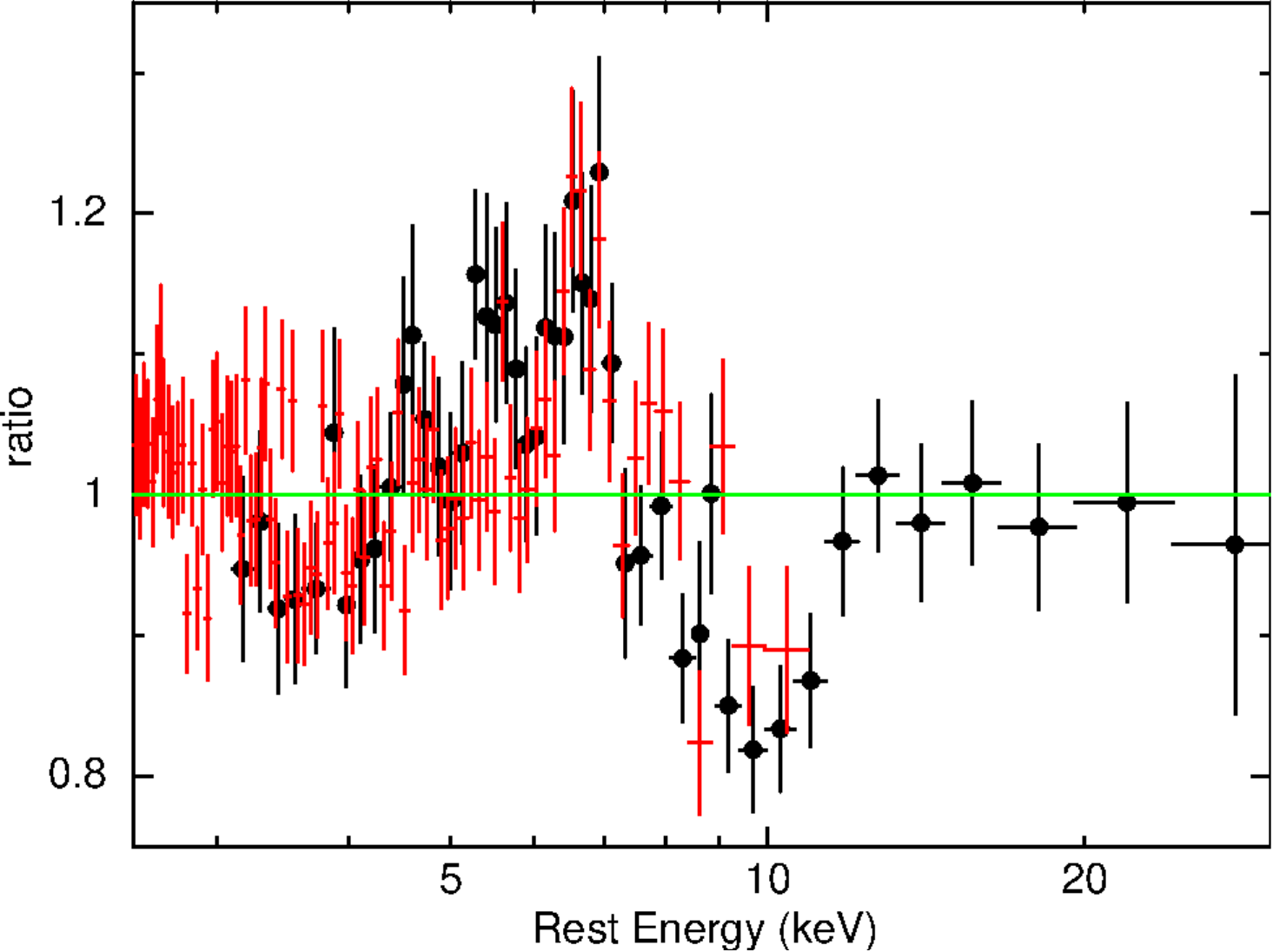} &
\includegraphics[width=0.42\textwidth,valign=t,clip]{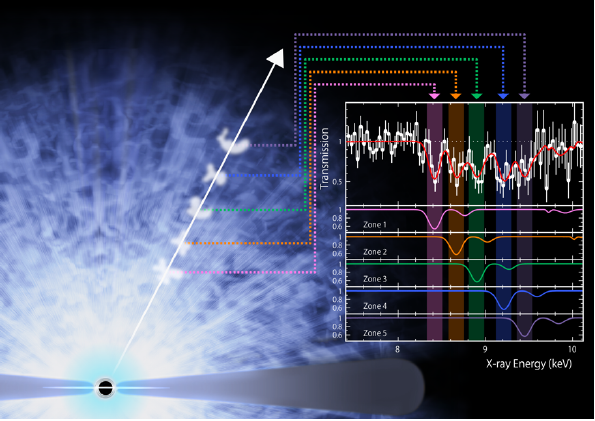}
\end{tabular}
\caption{
\textbf{Left:} Mean 2023 \textit{XMM-Newton/European Photon Imaging Camera pn} (EPIC-pn) and \textit{Nuclear Spectroscopic Telescope Array} (NuSTAR) spectra of PG\,1448+273 ($z$=0.0645), together with the data-to-continuum-model ratio. The broad absorption feature at $\sim$9--10\,keV in the AGN rest frame indicates a highly blueshifted Fe\,K absorber. Reproduced from \citet{Reeves2024}, licensed under CC BY 4.0.
\textbf{Right:} JAXA illustration of the UFO detected with \textit{XRISM/Resolve} in PDS\,456 ($z$=0.184; \citealt{XRISM2025PDS456}). The Fe\,K absorption complex is resolved into at least five highly ionised components with line-of-sight outflow velocities of $0.23$--$0.33c$. Courtesy of JAXA.
}
\label{fig:UFOspectra}
\end{figure}

Until recently, studies of highly ionised UFOs traced through Fe\,K absorption relied mainly on charge-coupled-device (CCD) spectroscopy, whose moderate spectral resolution generally prevented closely spaced velocity components from being resolved. The broad absorption feature in the near-Eddington Seyfert\,1 galaxy PG\,1448+273 illustrates this limitation (Fig.~\ref{fig:UFOspectra}, left panel). 
Launched in September 2023, \textit{XRISM} provides high-resolution Fe\,K spectroscopy with its \textit{Resolve} X-ray microcalorimeter. With the gate valve closed, \textit{Resolve} delivers a non-dispersive energy resolution of $\sim$5\,eV full width at half maximum (FWHM)  above $\sim$2\,keV, enabling individual velocity components of highly ionised Fe\,K UFOs to be distinguished for the first time, as demonstrated in PDS\,456 (Fig.~\ref{fig:UFOspectra}, right panel).\\

PDS\,456 is a well-studied AGN hosting powerful accretion-disc winds. At $z=$0.184, PDS\,456 is the most luminous radio-quiet AGN in the local Universe ($z$$<$0.3), with a bolometric luminosity of $\sim$10$^{47}$\,erg\,s$^{-1}$ and an Eddington ratio close to or slightly above unity, making it a nearby analogue of rapidly accreting quasars at cosmic noon ($z$$\sim$2--3).
Since the detection of its UFO in the 2001 \textit{XMM-Newton} observation \citep{Reeves2003}, repeated X-ray observations have revealed a recurrent, highly ionised outflow. Additional outflow signatures have been detected at UV, optical, IR and (sub-)mm wavelengths, from the nuclear region to kiloparsec scales. The first \textit{XRISM/Resolve} observations of PDS\,456 resolved the Fe\,K absorption into at least five velocity components, each detected with $>$99.99\% confidence \citep{XRISM2025PDS456}. Their line-of-sight outflow velocities span 0.23--0.33c, with similar column densities of 0.8--1.4$\times$10$^{23}$\,cm$^{-2}$ and ionisation parameters of $\log\xi$$\simeq$4.9. Interpreting these components as dense clumps embedded in a more diffuse ionised wind yields an inferred mass outflow rate of $\sim$60--300\,$M_\odot\,yr^{-1}$, compared with the earlier estimate of $\sim$10\,$M_\odot\,yr^{-1}$ inferred under a smooth-wind scenario \citep[e.g.][]{Nardini2015}. In this clumpy-wind scenario, the kinetic power reaches $\gtrsim$10$^{47}$$\,erg\,s^{-1}$, comparable to $L_{\rm bol}$, while the momentum flux is $\sim$10--50\,$L_{\rm bol}$/$c$. These energetics depend on the assumed wind geometry, gas density, radial location and volume filling factor.

Similarly, simultaneous \textit{XRISM} and \textit{XMM-Newton} observations provide a detailed view of the multiphase outflow in the near-Eddington AGN PG~1211+143 ($z$=0.0809). The \textit{Resolve} spectrum reveals several narrow Fe\,K absorption components, forming a ``UFO forest'' \citep{Mizumoto2026}, while the \textit{Reflection Grating Spectrometer} (RGS) onboard \textit{XMM-Newton} detects lower-ionisation UFO features at velocities similar to those of some of the Fe\,K components \citep{Reeves20262026PG1211}. This suggests that the highly ionised and lower-ionisation absorbers trace different phases of the same accretion-disc wind.\\

\textit{XRISM} also probes winds in less luminous AGN accreting at lower Eddington ratios. In NGC\,3783 ($z$=0.00973; $L_{\rm bol}$$\sim$2$\times$10$^{44}$\,erg\,s$^{-1}$; $L_{\rm bol}$/$L_{\rm Edd}$$\sim$0.06, where $L_{\rm Edd}$ is the Eddington luminosity), \textit{Resolve} detected six photoionised absorption components spanning a broad range of ionisation states, column densities and outflow velocities. These include a broad Fe\,{\sc xxvi} absorption component outflowing at $\sim$0.05c, with an inferred kinetic power of $\sim$0.8--3\% of $L_{\rm bol}$, within the range invoked in some AGN feedback models. The similarity between the Fe\,{\sc xxv} absorption profile and the UV Ly$\alpha$ and C\,{\sc iv} troughs indicates a clumpy, multi-zone outflow spanning a broad range of ionisation states \citep{Mehdipour2025}. A subsequent \textit{XRISM} observation detected a transient UFO at $\sim$0.19c during a bright X-ray/UV flare. \citet{Gu2025} interpreted this event as a magnetically driven ejection from the inner accretion disc, analogous to a solar coronal mass ejection. 
In NGC\,4051 ($z$=0.00234; $L_{\rm bol}$$\sim$3$\times$10$^{43}$\,erg\,s$^{-1}$; $L_{\rm bol}$/$L_{\rm Edd}$$\sim$0.25), \textit{Resolve} detected two variable blueshifted Fe\,K absorption components at outflow velocities of $\sim$0.025c and $\sim$0.04c  \citep{Reeves2026NGC4051}. The absorption components vary on timescales shorter than one day and shift towards higher velocities during the observation. This behaviour may arise from two separate absorbers transiting the line of sight or from a single accelerating outflow.  In the latter interpretation, the inferred acceleration favours magnetic over radiative driving. 
\textit{XRISM} has also provided a detailed view of the nuclear environment in NGC\,4151 ($z$=0.00332; $L_{\rm bol}$$\sim$5--8$\times$10$^{43}$\,erg\,s$^{-1}$; $L_{\rm bol}$/$L_{\rm Edd}$$\sim$0.01--0.02). \textit{Resolve} decomposed the Fe\,K$\alpha$ emission-line complex into three components associated, under a Keplerian-broadening interpretation, with the accretion disc, BLR and torus. It also detected up to six Fe\,K absorption components spanning different ionisation states and velocities, revealing a stratified, multiphase wind \citep{XRISM4151_2024,Miller2026}. Together, these observations constrain both the emitting and absorbing gas in the nuclear region.\\

ESA's \textit{NewAthena} mission \citep{Cruise2025}, currently planned for launch in 2039, will transform high-resolution X-ray studies of AGN winds by combining microcalorimeter spectroscopy with an order-of-magnitude increase in effective area relative to \textit{XRISM}, depending on energy. Its \textit{X-ray Integral Field Unit} (X-IFU) is designed to cover the $\sim$0.2--12\,keV band with a required energy resolution of $4$\,eV FWHM at 7\,keV ($3$\,eV goal; \citealt{Peille2025}). This combination of spectral resolution and sensitivity will enable wind kinematics, ionisation structure and variability to be characterised in much fainter nearby AGN and in luminous quasars at cosmic noon, providing tighter constraints on their mass outflow rates and kinetic powers. The \textit{Wide Field Imager} (WFI), combining high throughput with a $\sim$40$^{\prime}$$\times$40$^{\prime}$ field of view \citep{Antonelli2024}, will support systematic searches for UFO absorption in large AGN samples. Accounting for detection sensitivity and selection effects, these studies will constrain the incidence of UFOs and its dependence on luminosity, Eddington ratio and redshift. 
Together, \textit{X-IFU} and \textit{WFI} will establish how the physical properties and incidence of AGN winds evolve across black-hole growth and cosmic history.

\section{From accretion-disc winds to galaxy-scale outflows}
\label{sec:multi-wav}

Understanding how powerful X-ray winds launched from the inner accretion disc transfer their energy and momentum to the host galaxy remains a key open question in AGN feedback. While \textit{XRISM} is revealing the structure and energetics of UFOs on sub-parsec scales, linking them to the multiphase outflows observed over hundreds of parsecs to several kiloparsecs remains a major challenge. Addressing this connection requires complementary observations from the UV to the sub-mm domain to trace the different ionisation states and gas phases. Empirical relations between AGN luminosity and the energetics of ionised and molecular outflows support this multi-scale picture \citep{Fiore2017}.\\

UV spectroscopy probes gas that is generally less ionised than the highly ionised plasma observed in X-rays, revealing narrow absorption lines (NALs), mini-broad absorption lines (mini-BALs) and broad absorption lines (BALs), with velocities ranging from a few hundred to several tens of thousands of km\,s$^{-1}$. Together with WAs and UFOs, these UV absorbers may represent different phases of a stratified, multiphase accretion-disc wind \citep{Laha2021,Scepi2026}. In PG\,1211+143, simultaneous \textit{Hubble Space Telescope} (HST)/\textit{Cosmic Origins Spectrograph} (COS) and \textit{Chandra/High Energy Transmission Grating} (HETG) observations revealed a broad Ly$\alpha$ absorption trough at an outflow velocity of $\sim$0.056c, nearly identical to that of the contemporaneous low-ionisation X-ray UFO \citep{Danehkar2018,Kriss2018}. This velocity correspondence suggests that the UV and X-ray absorbers trace different phases of the same accretion-disc wind, possibly with dense, lower-ionisation clumps embedded in a more highly ionised outflow \citep{Pounds2016,Reeves2018}. 
A related case is PDS~456, where a 2000 \textit{HST}/Space Telescope Imaging Spectrograph (STIS) spectrum revealed a broad UV absorption trough, originally interpreted as Ly$\alpha$ at $v_{\rm out}$$\sim$0.06c  \citep{OBrien2005}. \citet{Hamann2018} instead identified the feature as C\,{\sc iv} at $v_{\rm out}$$\sim$0.30c, comparable to the X-ray UFO velocity. If this latter identification is confirmed, it would provide further evidence for a lower-ionisation phase associated with the highly ionised accretion-disc wind. The long-lived obscuration event in NGC\,5548 also demonstrated that a clumpy, weakly ionised outflow located near the BLR can obscure up to $\sim$90\% of the soft X-ray emission \citep{Kaastra2014,Mehdipour2015}. Together, these observations demonstrate the importance of combining UV and X-ray spectroscopy to determine the multiphase structure of AGN accretion-disc winds.\\

At larger scales, optical, IR and (sub-)mm observations probe the ionised, warm molecular and cold molecular phases of AGN-driven outflows \citep[e.g.][]{Harrison2018}. PDS\,456 currently provides one of the most complete multi-wavelength views of such an outflow. \textit{Atacama Large Millimeter/submillimeter Array (ALMA)} observations reveal a clumpy CO(3--2) molecular outflow extending to $\sim$5\,kpc, with a mass outflow rate of $\sim$290\,$M_\odot\,yr^{-1}$ \citep{Bischetti2019}. \textit{Multi Unit Spectroscopic Explorer} (MUSE)  observations reveal a large-scale [O\,{\sc iii}] outflow extending to $\sim$12\,kpc, while the H$\alpha$ outflow is detected within $\sim$3\,kpc and exhibits a morphology and kinematics similar to those of the CO(3--2) outflow \citep{Travascio2024}. Combined \textit{James Webb Space Telescope} (JWST) and \textit{MUSE} observations trace a multiphase outflow to $\sim$15\,kpc through Pa$\alpha$, warm H$_2$ and ionised-gas lines including [Ne\,{\sc iii}] and [Ne\,{\sc vi}] \citep{Seebeck2024}. These observations demonstrate that the outflow remains multiphase over kiloparsec scales and provide the measurements required to constrain how mass, momentum and energy propagate through the host galaxy.\\

The revised nuclear-wind energetics inferred from the \textit{XRISM} observations of PDS\,456 have important implications for this connection. Under the clumpy-wind interpretation, the instantaneous momentum flux of the accretion-disc wind is approximately an order of magnitude larger than that of the galaxy-scale outflow, while its kinetic power is more than three orders of magnitude larger \citep{XRISM2025PDS456}. These differences disfavour the simplest steady momentum- and energy-conserving coupling scenarios \citep{Faucher2012}. 
However, a direct comparison of the instantaneous nuclear-wind energetics with those of galaxy-scale outflows, which trace AGN activity over much longer timescales, is intrinsically uncertain. Both estimates also remain sensitive to assumptions about geometry, density and gas distribution \citep{Bischetti2019,Travascio2024,XRISM2025PDS456}. 
The observed differences may therefore reflect an intermittent AGN duty cycle or inefficient transfer of energy and momentum from the nuclear wind to the interstellar medium \citep{XRISM2025PDS456}. Coordinated X-ray, optical, IR and (sub-)mm observations of larger AGN samples will be required to determine whether the extreme nuclear-wind energetics inferred for PDS\,456 are representative and how efficiently accretion-disc winds drive galaxy-scale outflows.

\section{Conclusion}\label{sec:conclusion}

AGN winds and outflows are intrinsically multi-scale, multiphase and time-variable phenomena, extending from the immediate vicinity of the SMBH to galaxy-wide scales.  No single tracer can provide a complete view of these outflows. Coordinated observations from X-rays to the sub-mm domain are required to determine how mass, momentum and energy propagate through their highly ionised, ionised, atomic and molecular phases.

High-resolution spectroscopy with X-ray microcalorimeters has opened a new era in the study of accretion-disc winds. The first \textit{XRISM/Resolve} observations have revealed that UFOs are substantially more kinematically structured and inhomogeneous than inferred from CCD spectroscopy. In PDS\,456, resolving the Fe\,K absorption into multiple velocity components has led to a substantial upward revision of the inferred mass outflow rate, momentum flux and kinetic power. 
These results strengthen the case that accretion-disc winds can provide the energy budget required for AGN feedback, while identifying their coupling efficiency and duty cycle as key factors governing their impact on the host galaxy. 

Establishing this connection will require the complementary capabilities of current and future multi-wavelength facilities. 
Building on the discoveries of \textit{XRISM}, ESA’s \textit{NewAthena} mission will transform the study of AGN winds by combining detailed \textit{X-IFU} spectroscopy of individual sources with \textit{WFI} surveys of large AGN samples, revealing how wind properties, occurrence rates and duty cycles evolve from faint nearby AGN to luminous quasars at cosmic noon. 
Together with observations from facilities such as the \textit{ELT}, \textit{JWST}, \textit{ALMA} and \textit{SKA}, these measurements will connect nuclear winds with the ionised, atomic and molecular gas on galactic scales. They will ultimately determine whether and with what efficiency  accretion-disc winds contribute to the co-evolution of SMBHs and their host galaxies.

\begin{acknowledgements}
\section*{Acknowledgements}
I warmly thank the organisers of Session 20 for their kind invitation to present this review on AGN winds and outflows, and all the organisers of the SF2A 2026 annual meeting for making it such an enjoyable and stimulating  scientific event. 
\end{acknowledgements}

\small
\bibliographystyle{aa}  
\bibliography{Porquet_S20} 

\end{document}